\documentclass[showpacs,preprintnumbers,amsmath,amssymb,prb,twocolumn]{revtex4}

\def\XXint#1#2#3{{\setbox0=\hbox{$#1{#2#3}{\int}$}
     \vcenter{\hbox{$#2#3$}}\kern-.5\wd0}}

\usepackage{graphicx}
\usepackage{subfigure}
\usepackage{bm}

\begin{document}

\title{Formation of spiral ordering by magnetic field in frustrated anisotropic antiferromagnets}

\author{O.\ I.\ Utesov$^{1,2,3}$}
\email{utiosov@gmail.com}
\author{A.\ V.\ Syromyatnikov$^{1,3}$}
\email{asyromyatnikov@yandex.ru}

\affiliation{$^1$National Research Center ``Kurchatov Institute'' B.P.\ Konstantinov Petersburg Nuclear Physics Institute, Gatchina 188300, Russia}
\affiliation{$^2$St. Petersburg Academic University - Nanotechnology Research and Education Centre of the Russian Academy of Sciences, St.\ Petersburg 194021, Russia}
\affiliation{$^3$St.\ Petersburg State University, 7/9 Universitetskaya nab., St.\ Petersburg 199034, Russia}

\date{\today}

\begin{abstract}

We discuss theoretically phase transitions in frustrated antiferromagnets with biaxial anisotropy or dipolar forces in magnetic field applied along the easy axis at $T=0$. There are well-known sequences of phase transitions upon the field increasing: the conventional spin-flop transition and the flop of the spiral plane at strong and weak easy-axis anisotropy, respectively. We argue that much less studied scenarios can appear at moderate anisotropy in which the magnetic field induces transitions of the first order from the collinear state to phases with spiral orderings. Critical fields of these transitions are derived in the mean-field approximation and the necessary conditions are found for the realization of these scenarios. We show that one of the considered sequences of phase transitions was found in multiferroic MnWO$_4$ both experimentally and numerically (in a relevant model) and our theory reproduces quantitatively the numerical findings.

\end{abstract}


\maketitle

\section{Introduction}
\label{Intro}

Frustrated antiferromagnets (AFs) attract significant attention now due to their rich phase diagrams and multiferroic properties of some of their phases with non-collinear magnetic ordering (see, e.g., Refs~\cite{nagaosa,Ehrenberg1997,Kimura2003,mni3,utesov2017}). Multiferroics of spin origin in which ferroelectricity is induced by spiral magnetic order show a giant magnetoelectric response (see, e.g., Refs~\cite{Cheong2007,Tokura2009}) that makes them promising materials for technological applications. The frustration plays an important role in such multiferroics providing the non-collinear spin textures. For example, non-collinear magnetic phases in frustrated magnet MnWO$_4$ were shown to be ferroelectric. \cite{arkenbout2006,nojiri2011,mitamura2012} Thus, phase transitions in frustrated AFs governed by external magnetic field is an important topic now.

\begin{figure}
  \centering
  \includegraphics[width=8cm]{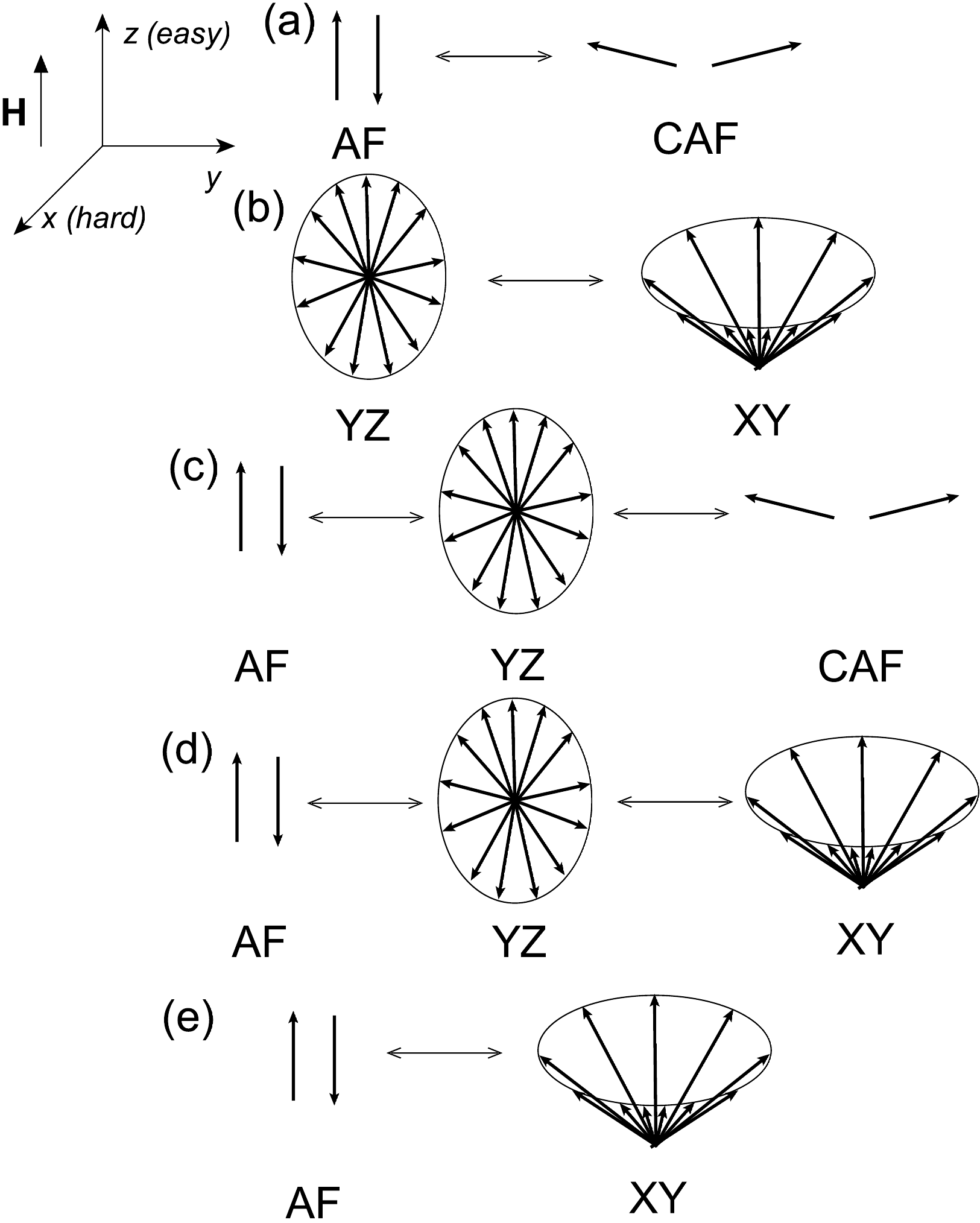}\\
  \caption{Possible scenarios of phase transitions in frustrated antiferromagnet with Hamiltonian~\eqref{ham1} upon magnetic field $\bf H$ increasing when $\bf H$ is directed along the easy axis $z$. (a) Strong easy-axis anisotropy. The conventional spin-flop transition in which the collinear antiferromagnetic (AF) phase is followed by the canted AF phase (CAF) at $H>H_{sf}$. (b) Weak anisotropy. First-order transition from the spiral phase (YZ) in which all spins lie in the easy $yz$ plane to the conical spiral phase (XY) in which spins rotate in $xy$ plane.
(c)--(e) Scenarios discussed in the present paper which are expected to arise at moderate anisotropy.
\label{fig1}}
\end{figure}

The plane in which spins rotate (spiral plane) is selected in real materials by small anisotropic spin interactions. As a result, application of small or moderate magnetic fields within the spiral plane produces a flop of the spiral plane in many multiferroics accompanied with the flop of the electric moment. \cite{nagaosa} We address this effect in frustrated AFs with small biaxial anisotropy in our previous paper \cite{utesov2018} and show that the flop of the spiral plane resembles the conventional spin-flop transition in collinear AFs \cite{Neel1936} (see Figs.~\ref{fig1}(a) and \ref{fig1}(b)). Critical fields at which these transitions take place are given by similar formulas having the structure $S \sqrt{DJ}$, where $J$ is the characteristic energy of the exchange interaction, $D \ll J$ is the anisotropy value, and $S$ is the spin value.

In the present paper, we continue the discussion of anisotropic frustrated AFs in small or moderate magnetic fields and consider evolution of phase transitions upon variation of the anisotropy value in a simple model containing the frustrated exchange interaction and the single-ion biaxial anisotropy (or dipolar forces). Applying the field along the easy direction, we observe the conventional spin-flop transition presented in Fig.~\ref{fig1}(a) at sufficiently strong easy-axis anisotropy. At weak anisotropy, we find the spiral plane flop shown in Fig.~\ref{fig1}(b) which was discussed in detail in our previous paper \cite{utesov2018}. The main goal of the present study is quantitative consideration of the moderate anisotropy regime in the mean-field approximation. We propose novel sequences of phase transitions presented in Figs.~\ref{fig1}(c)--(e). Scenario shown in Figs.~\ref{fig1}(c) can be interpreted as the spin-flop transition splitting into two first-order transitions with an intermediate spiral phase. In Sec.~\ref{Frust} we find expressions for the critical fields and conditions for realization of these scenarios of phase transitions.

In Sec.~\ref{applic}, we present some particular sets of model parameters at which scenarios shown in Figs.~\ref{fig1}(c)--(e) arise. We demonstrate that the scenario of phase transitions depicted in Fig.~\ref{fig1}(c) is realized in the  considered model with parameters proposed in Ref.~\cite{zh} for description of experimentally obtained phase diagram of MnWO$_4$. We present a summary of results and our conclusions in Sec.~\ref{Conc}.

\section{Frustrated antiferromagnets. General consideration.}
\label{Frust}

In this section, we present a general consideration of simple models in which a subtle interplay between different magnetic interactions leads to sequences of phase transitions shown in Fig.~\ref{fig1}.

\subsection{Antiferromagnets with single-ion biaxial anisotropy}
\label{Theor}

We consider the frustrated Heisenberg AF with small single-ion biaxial anisotropy whose Hamiltonian has the form
\begin{eqnarray}
 \label{ham1}
  \mathcal{H} &=& \mathcal{H}_{ex} + \mathcal{H}_{an} + \mathcal{H}_{z}, \nonumber \\
  \mathcal{H}_{ex} &=& -\frac12 \sum_{i,j} J_{ij} \left(\mathbf{S}_i \cdot \mathbf{S}_j\right), \\
  \mathcal{H}_{an} &=& - \sum_i \left[ D(S_i^z)^2 + E (S_i^y)^2\right], \nonumber \\
  \mathcal{H}_z &=& - \sum_i \left(\mathbf{h} \cdot \mathbf{S}_i\right),\nonumber
\end{eqnarray}
where ${\bf h}=g \mu_B {\bf H}$ is the magnetic field in energy units and we assume for definiteness that $D > E > 0$ so that $x$ and $z$ are the hard and the easy axes, respectively. We also assume in all general derivations below that there is one spin in a unit cell and the lattice is arbitrary. After the Fourier transform
\begin{equation}
\label{four1}
  \mathbf{S}_j = \frac{1}{\sqrt{N}} \sum_\mathbf{q} \mathbf{S}_\mathbf{q} e^{i \mathbf{q} \mathbf{R}_j},
\end{equation}
where $N$ is the number of spins in the lattice, Hamiltonian \eqref{ham1} acquires the form
\begin{eqnarray}
  \label{ex2}
  \mathcal{H}_{ex} &=& -\frac12 \sum_\mathbf{q} J_\mathbf{q} \left(\mathbf{S}_\mathbf{q} \cdot \mathbf{S}_{-\mathbf{q}}\right), \\
	\label{an21}
  \mathcal{H}_{an} &=& - \sum_\mathbf{q}\left[ D S^z_\mathbf{q} S^z_{-\mathbf{q}} + E S^y_\mathbf{q} S^y_{-\mathbf{q}}\right], \\
	\label{z21}
 \mathcal{H}_z &=& - \sqrt{N} \left(\mathbf{h} \cdot \mathbf{S}_{\bf 0}\right).
\end{eqnarray}
We assume that $J_\mathbf{q}$ has two equivalent maxima at ${\bf q}=\pm {\mathbf{k}}$. Then, in the absence of the anisotropy, the ground state of the system at $h=0$ is a plane spiral with modulation vector $\mathbf{k}$. We consider a simple case in which strong enough anisotropy leads to a collinear antiferromagnetic (AF) structure characterized by the vector ${\bf q}=\mathbf{k}_0$ in which spins are directed along $z$ axis at $h=0$ and the average magnetization is zero ($\mathbf{k}_0$ can be equal to, e.g., $(\pi,\pi,\pi)$, $(0,0,\pi)$, etc.). In general, there can be also other more complicated collinear structures one of which is discussed in Sec.~\ref{Onefourth}.

At finite $\bf h$ applied along $z$ axis, the competing spin structures are the following (see Fig.~\ref{fig1}): (i) the collinear AF phase, (ii) the canted AF state (CAF), (iii) the helical state in which spins rotate in the easy $yz$ plane (YZ), and (iv) the conical spiral in which spins rotate in the $xy$ plane (XY). Due to the anisotropy, the AF state has lower energy than CAF at small $h$ and YZ has lower energy than XY. Classical ground state energies $\cal E$ of the considered structures read as
\begin{equation}
\label{enaf}
  \frac{1}{N} \mathcal{E}_{AF} = -\frac{S^2 }{2} J_{\mathbf{k}_0}-S^2D,
\end{equation}
\begin{equation}
\label{ensf}
  \frac{1}{N} \mathcal{E}_{CAF} \approx -\frac{S^2 }{2} J_{\mathbf{k}_0}- S^2 E - \frac{h^2}{2(J_{\mathbf{k}_0}-J_{0})},
\end{equation}
\begin{equation}
\label{enyz}
  \frac{1}{N} \mathcal{E}_{YZ} \approx -\frac{S^2 }{2} J_\mathbf{k}- \frac{S^2(D+E)}{2}- \frac{h^2}{2(2 J_{\mathbf{k}}-J_{0}-J_{2\mathbf{k}})},
\end{equation}
\begin{equation}
\label{enxy}
  \frac{1}{N} \mathcal{E}_{XY} \approx -\frac{S^2 }{2} J_{\mathbf{k}}- \frac{S^2 E }{2} - \frac{h^2}{2(J_{\mathbf{k}}-J_{0})}.
\end{equation}
The detailed derivation of Eqs.~\eqref{enyz} and \eqref{enxy} can be found in Ref.~\cite{utesov2018}. Eqs.~\eqref{enyz} and \eqref{enxy} are obtained in the first order in $D$ and $E$, under assumption that $h$ is of the order of the conventional spin-flop field
\begin{equation}
\label{hsf}
	h_{sf} = S \sqrt{2(D-E)(J_{\mathbf{k}_0}-J_{0})}
\end{equation}
which is much smaller than the saturation field $h_s\sim SJ$. Eq.~\eqref{hsf} is found by comparing Eqs.~\eqref{enaf} and \eqref{ensf}. We also neglect higher order harmonics in spiral phases which arise due to the anisotropy. As it was shown in Ref.~\cite{utesov2018}, contributions from higher harmonics to the ground state energy read as (notice, they are of the second order in the anisotropy)
\begin{equation}\label{harm1}
  -\frac{S^2(D-E)^2}{2(J_\mathbf{k}-J_{3\mathbf{k}})}\, \text{ and } \, -\frac{S^2 E^2}{2(J_\mathbf{k}-J_{3\mathbf{k}})}
\end{equation}
for YZ and XY structures, respectively. Thus, our approach is valid if
\begin{eqnarray}
\label{cond1}
 D-E &\ll& \min\{ J_{\mathbf{k}}-J_{3\mathbf{k}}, J_{\mathbf{k}_0}-J_{0} \},\nonumber\\
 E &\ll& \min\{ J_{\mathbf{k}}-J_{3\mathbf{k}}, J_{\mathbf{k}_0}-J_{0} \}
\end{eqnarray}
(see Ref.~\cite{utesov2018} for more details).

One can see from Eqs.~\eqref{enaf}--\eqref{enxy} that the AF phase is stable at $h=0$ if
\begin{equation}
\label{cond2}
  D-E > J_{\mathbf{k}}-J_{\mathbf{k}_0} \equiv \alpha.
\end{equation}
Besides, the CAF phase is energetically preferable in comparison with XY one if
\begin{equation}
\label{cond3}
	E > \alpha.
\end{equation}
 The opposite case of $D-E<\alpha$ and $E<\alpha$ is considered in detail in Ref.~\cite{utesov2018}, where the spiral plane flop was observed upon the field increasing (i.e., the transition shown in Fig.~\ref{fig1}(b)). Conditions \eqref{cond1} and \eqref{cond2} are compatible with each other if $\mathbf{k}$ is not very close to and not very far from $\mathbf{k}_0$ (we also imply here that $2\mathbf{k}_0$ is equal to a reciprocal lattice vector as it is frequently the case in AF phases). As it is shown in Sec.~\ref{applic}, this can be achieved in a rather broad range of model parameters.

If conditions \eqref{cond1}--\eqref{cond3} hold, one has AF$\leftrightarrow$YZ$\leftrightarrow$CAF sequence of phase transitions instead of the conventional scenarios of AF$\leftrightarrow$CAF and YZ$\leftrightarrow$XY. The critical field at which the AF$\leftrightarrow$YZ transition takes place can be found from Eqs.~\eqref{enaf} and \eqref{enyz}, the result being
\begin{equation}
\label{h1}
  h_1 = S \sqrt{(D-E-\alpha)(2 J_{\mathbf{k}}-J_{0}-J_{2\mathbf{k}})}.
\end{equation}
The critical field of YZ$\leftrightarrow$CAF transition derived from Eqs.~\eqref{ensf} and \eqref{enyz} has the form
\begin{equation}
\label{h2}
  h_2 = S \sqrt{(D-E+\alpha) \frac{(2 J_{\mathbf{k}}-J_{0}-J_{2\mathbf{k}})(J_{\mathbf{k}_0}-J_{0})}{2 J_{\mathbf{k}}-J_{\mathbf{k}_0}-J_{2\mathbf{k}}}}.
\end{equation}
The condition of existence of YZ phase, $h_1<h_2$, reads as
\begin{equation}
\label{exyz}
  \alpha < D-E < \alpha \frac{2 J_{\mathbf{k}}-J_{0}-J_{2\mathbf{k}}}{2 J_{\mathbf{k}}-J_0-J_{2\mathbf{k}}-2(J_{\mathbf{k}_0}-J_0)},
\end{equation}
where we take into account also Eq.~\eqref{cond2} and assume that the denominator is positive. Bearing in mind the positiveness of $J_{\mathbf{k}_0}-J_0$, one concludes that Eq.~\eqref{exyz} gives a finite interval for $D-E$. If the denominator in Eq.~\eqref{exyz} is negative, $h_1<h_2$ if Eqs.~\eqref{cond1} and \eqref{cond2} holds.

One can see from Eqs.~\eqref{ensf} and \eqref{enxy} that XY phase is energetically preferable in comparison with CAF state if
\begin{equation}
\label{cond4}
	E<\alpha.
\end{equation}
In this case, two possible sequences of phase transitions can appear which are presented in Fig.~\ref{fig1}(d) and \ref{fig1}(e). The first one is AF$\leftrightarrow$YZ$\leftrightarrow$XY. The field of AF$\leftrightarrow$YZ transition is given by Eq.~\eqref{h1}. YZ$\leftrightarrow$XY transition is of the spiral plane flop type which is described in detail in Ref.~\cite{utesov2018} and which arises at $h=h_{sp}$, where
\begin{equation}
\label{hsp}
  h_{sp} = S \sqrt{D \frac{(2 J_{\mathbf{k}}-J_{0}-J_{2\mathbf{k}})(J_{\mathbf{k}}-J_{0})}{J_{\mathbf{k}}-J_{2\mathbf{k}}}}.
\end{equation}
This scenario appears if
\begin{equation}
\label{cond5}
	J_0 \leq J_{2\mathbf{k}}
	\quad\mbox{ or }\quad
	D < (E + \alpha) \frac{J_{\mathbf{k}}-J_{2\mathbf{k}}}{J_0-J_{2\mathbf{k}}}.
\end{equation}
When both of these conditions are violated, one has
 \begin{equation}
\label{cond6}
	 h_1>h_{sp},
 \end{equation}
where $h_1$ and $h_{sp}$ are given by Eqs.~\eqref{h1} and \eqref{hsp}, respectively, and the sequence of phase transitions shown in Fig.~\ref{fig1}(e) (AF$\leftrightarrow$XY) takes place. Corresponding critical field derived from Eqs.~\eqref{enaf} and~\eqref{enxy} reads as
\begin{equation}
\label{hxy}
  h_{xy} = S \sqrt{(2D-E-\alpha) (J_{\mathbf{k}}-J_{0})}.
\end{equation}

\subsection{Antiferromagnets with dipolar forces}
\label{Dip}

In low-symmetry lattices, the magneto-dipolar interaction can effectively produce the biaxial anisotropy. \cite{utesov2017,utesov2018} Moreover, dipolar forces can be the main source of anisotropy in systems containing magnetic ions with half-filled $d$-shells (e.g., Mn$^{2+}$) because the spin-orbit interaction is particularly small in them. Then, we consider in this subsection the model with Hamiltonian \eqref{ham1} in which $\mathcal{H}_{an}$ is replaced by
\begin{eqnarray}
 \label{ham4}
  \mathcal{H}_d &=& \frac12 \sum_{i,j} D^{\alpha \beta}_{ij} S^\alpha_i S^\beta_j, \\
	 {\cal D}^{\alpha \beta}_{ij} &=& \omega_0 \frac{v_0}{4 \pi} \left( \frac{1}{R_{ij}^3} - \frac{3 R_{ij}^\alpha R_{ij}^\beta }{R_{ij}^5}\right), \nonumber
\end{eqnarray}
where $v_0$ is the unit cell volume and
\begin{equation}\label{dipen}
  \omega_0 = 4 \pi \frac{(g \mu_B)^2}{v_0} \ll J
\end{equation}
is the characteristic dipolar energy. After Fourier transform \eqref{four1} we have
\begin{equation}\label{dip2}
  \mathcal{H}_d = \frac12 \sum_\mathbf{q} {\cal D}^{\alpha \beta}_\mathbf{q} S^\alpha_\mathbf{q} S^\beta_{-\mathbf{q}}.
\end{equation}
Tensor ${\cal D}^{\alpha \beta}_\mathbf{q}/2$ has three eigenvalues $\lambda_1(\mathbf{q}) \geq \lambda_2(\mathbf{q}) \geq \lambda_3(\mathbf{q})$ corresponding to three orthogonal eigenvectors $\mathbf{v}_1(\mathbf{q})$, $\mathbf{v}_2(\mathbf{q})$, and $\mathbf{v}_3(\mathbf{q})$. There is a correspondence with the model having the single-ion biaxial anisotropy if we denote $D=\lambda_1(\mathbf{q}) -\lambda_3(\mathbf{q})$  and $E=\lambda_1(\mathbf{q}) -\lambda_3(\mathbf{q})$ and direct $z$ axis along $\mathbf{v}_3(\mathbf{q})$, $y$ axis along $\mathbf{v}_2(\mathbf{q})$, and $x$ axis along $\mathbf{v}_1(\mathbf{q})$. The spiral vector $\mathbf{k}$ minimizes $-J(\mathbf{q})+ [ \lambda_2(\mathbf{q}) + \lambda_3(\mathbf{q})]/2$ and it is close to the momentum which maximizes $J(\mathbf{q})$.

If $\lambda_i(\mathbf{k}) \approx \lambda_i(\mathbf{k}_0)$ and $\mathbf{v}_i(\mathbf{k}) \approx \mathbf{v}_i(\mathbf{k}_0)$ at $i=1,2,3$, results of Sec.~\ref{Theor} are directly applicable to the considered situation upon the substitutions $D \rightarrow \lambda_1(\mathbf{q}) -\lambda_3(\mathbf{q})$ and $E \rightarrow \lambda_1(\mathbf{q}) -\lambda_2(\mathbf{q})$. However, the eigenvalues and eigenvectors can differ at momenta $\mathbf{k}_0$ and $\mathbf{k}$. Thus, easy and hard axes can be different in the collinear and the spiral structures. This complicates the behaviour of the system under external magnetic field. Corresponding analysis is out of the scope of the present paper.

\section{Frustrated antiferromagnets. Applications.}
\label{applic}

\subsection{Chain of classical spins}
\label{Model}

\begin{figure}
  \centering
  \includegraphics[width=8cm]{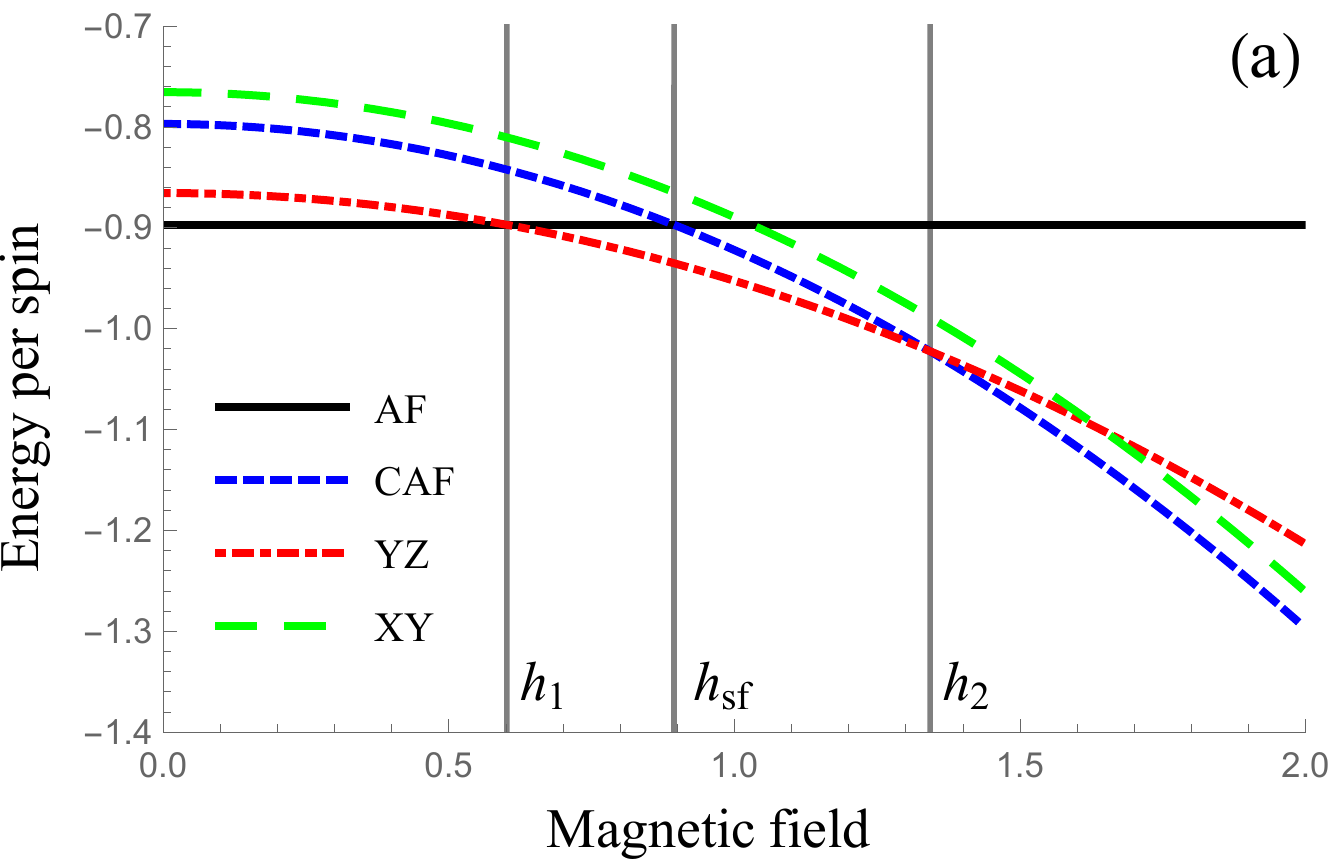}
	\includegraphics[width=8cm]{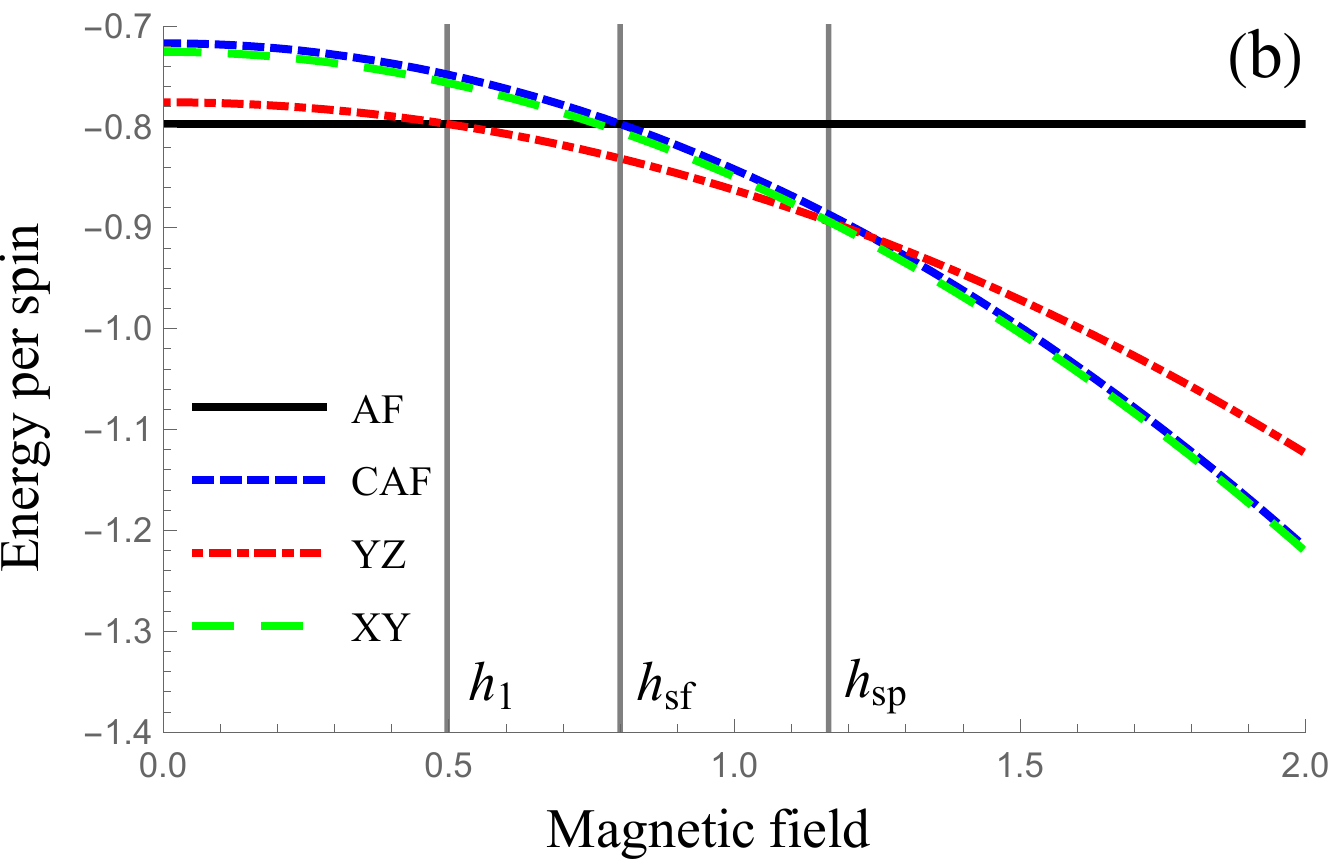}
	\includegraphics[width=8cm]{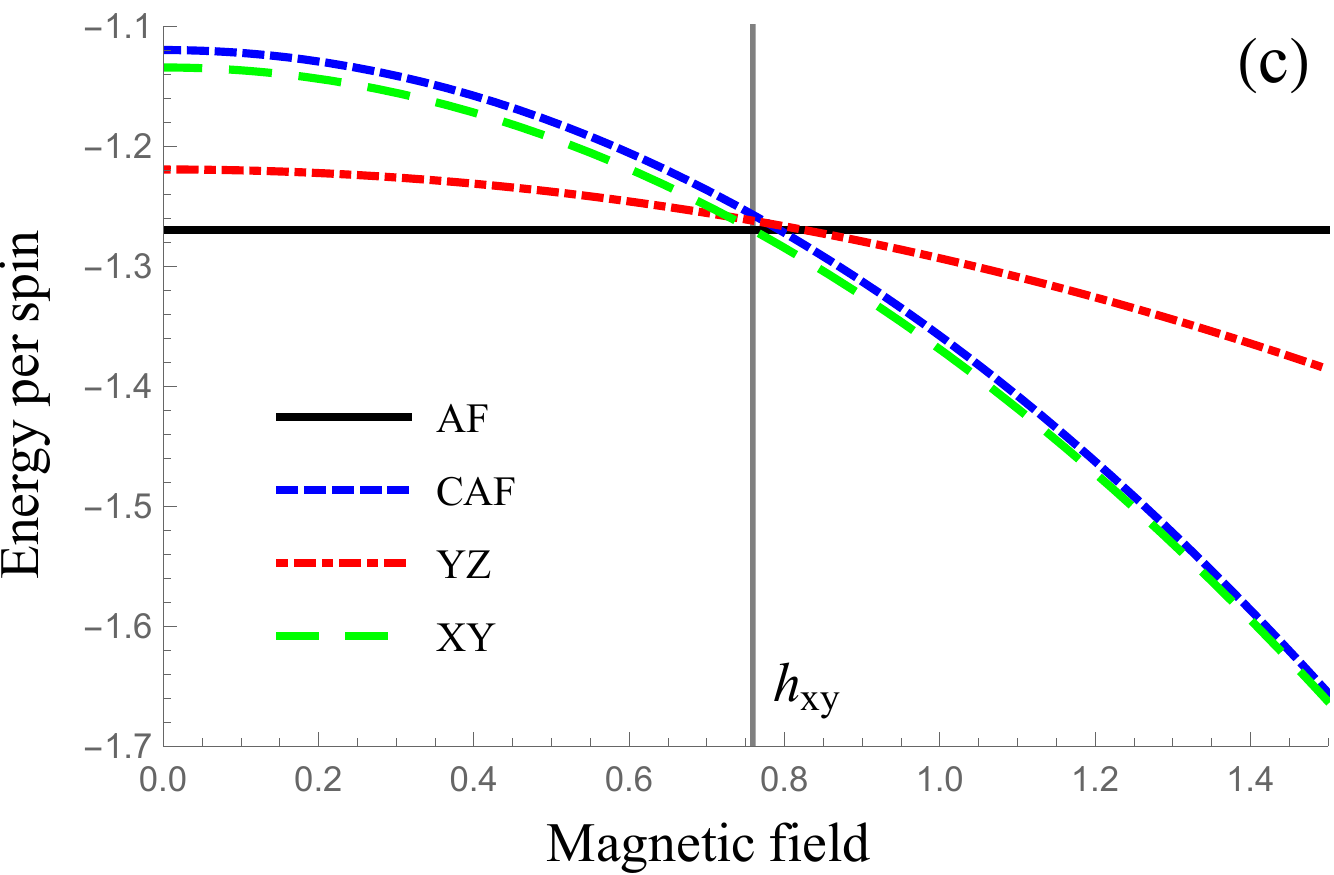}
  \caption{Ground state energies of competing phases \eqref{enaf}--\eqref{enxy} for the set of parameters (a) \eqref{Apar1}, (b) \eqref{Apar2}, and (c) \eqref{Apar3}. Critical fields $h_1$, $h_2$, $h_{sp}$, and $h_{xy}$ as well as $h_{sf}$ are denoted by gray vertical lines which are given by Eqs.~\eqref{h1}, \eqref{h2}, \eqref{hsp}, \eqref{hxy}, and \eqref{hsf}, respectively.
	\label{plot1}}
\end{figure}

We discuss now a particular realization of model \eqref{ham1} in which considered sequences of phase transitions can arise: a system of classical spin chains with two competing antiferromagnetic exchange interactions $J_1$ and $J_2$ between nearest and next-nearest spins. This model is relevant to 3D spin systems containing ferromagnetic planes interacting with each other by the frustrating AF interactions. Each ferromagnetic plane plays a role of a classical magnetic moment in the mean-field consideration of the spin ordering.
We have in this case
\begin{equation}\label{AJq}
  J_\mathbf{q} = - 2 (J_1 \cos{q_c} + J_2 \cos{2q_c}).
\end{equation}
If $J_2  > J_1/4$, $J_\mathbf{q}$ has a maximum at $\mathbf{k}=(0,0,k)$, where
\begin{equation}\label{Ak}
  k= \pi - \arccos{\frac{J_1}{4J_2}}.
\end{equation}
Let us consider the following set of dimensionless parameters
\begin{equation}
\label{Apar1}
  J_1=1, \quad J_2 = 0.3, \quad D=0.2, \quad E=0.1, \quad S=1
\end{equation}
which gives $k\approx 0.81 \pi$, $\mathbf{k}_0=(0,0,\pi)$, $J_{\mathbf{k}}-J_{\mathbf{k}_0} \approx 0.04$, and $J_{\mathbf{k}}-J_{3\mathbf{k}} \approx 1.35$. Then, conditions \eqref{cond1}, \eqref{cond2}, and \eqref{cond3} are well satisfied and the scenario shown in Fig.~\ref{fig1}(c) is realized. Eqs.~\eqref{h1} and \eqref{h2} give $h_1 \approx 0.6$ and $h_2 \approx 1.34$. Field $h_{sf} \approx 0.9$ given by Eq.~\eqref{hsf} lies in between of $h_1$ and $h_2$. Ground state energies \eqref{enaf}--\eqref{enxy} of considered spin states are drawn in Fig.~\ref{plot1}(a). Notice that XY conical spiral has higher energy than CAF. The saturation field, which can be estimated as $h_s \approx S(J_\mathbf{k}-J_0) \approx 4$ is not shown in Fig.~\ref{plot1}. One can replace $J_2$ in Eq.~\eqref{Apar1} by any value from the interval $(0.27,0.34)$ to realize the considered scenario of phase transitions AF$\leftrightarrow$YZ$\leftrightarrow$CAF.

The sequence of phase transitions AF$\leftrightarrow$YZ$\leftrightarrow$XY (see Fig.~\ref{fig1}(d)) appears with the following set of parameters:
\begin{equation}
\label{Apar2}
  J_1=1, \quad J_2 = 0.3, \quad D=0.1, \quad E=0.02, \quad S=1.
\end{equation}
Evidently, conditions \eqref{cond4} and \eqref{cond5} ($J_{0}<J_{2\mathbf{k}}$) hold in this case. Eqs.~\eqref{h1} and~\eqref{hsp} yields $h_1 \approx 0.5$ and $h_{sp} \approx 1.17 $, respectively. Corresponding ground state energies are plotted in Fig.~\ref{plot1}(b). One can replace $J_2$ in Eq.~\eqref{Apar2} by any value from the interval $(0.29,0.33)$ to realize this scenario of phase transitions.

The scenario depicted in Fig.~\ref{fig1}(e) (AF$\leftrightarrow$XY) can appear if one includes the third-nearest-neighbor exchange interaction along the chain so that
\begin{equation}\label{AJq2}
  J_\mathbf{q} = - 2 (J_1 \cos{q_c} + J_2 \cos{2q_c} + J_3\cos{3q_c}).
\end{equation}
Exchange constants
\begin{equation}\label{Apar3}
  J_1=1, \quad J_2 = -0.5, \quad J_3 = -0.4,
\end{equation}
give $k\approx 0.87 \pi$, $J_{\mathbf{k}}-J_{\mathbf{k}_0} \approx 0.05$, $J_{0}-J_{2\mathbf{k}} \approx 1.86$, and $J_{\mathbf{k}}-J_{3\mathbf{k}} \approx 1.69$. For this set of parameters, the scenario AF$\leftrightarrow$XY appears if $E<\alpha$ and $D>0.16$ (see Eqs.~\eqref{cond4} and \eqref{cond5}). Ground state energies are plotted in Fig.~\ref{plot1}(c) for
\begin{equation}
\label{Apar4}
  D=0.17, \quad E=0.02
\end{equation}
which satisfy these conditions. Eq.~\eqref{hxy} gives $h_{xy} \approx 0.76$.

To confirm analytical results presented in this subsection, we perform numerical simulations using Monte-Carlo method on chains containing $500$ and $1000$ sites. We use $10^6$ and $2 \cdot 10^6$ numbers of steps. The results obtained are almost independent of these numbers of sites and steps. Numerical calculations quantitatively reproduce with high accuracy spin orderings and ground state energies depicted in Fig.~\ref{plot1} for all considered sets of parameters. In particular, we get for parameters~\eqref{Apar1} $h_1 \approx 0.61$, $h_2 \approx 1.24$, and $h_{sf} \approx 0.87$ (cf.\ Fig.~\ref{plot1}(a)). For parameters~\eqref{Apar2}, we obtain $h_1 \approx 0.52$ and $h_{sp} \approx 1.07$ (cf.\ Fig.~\ref{plot1}(b)). For parameters listed in Eqs.~\eqref{Apar3} and~\eqref{Apar4}, we get $h_{xy} \approx 0.77$ (cf.\ Fig.~\ref{plot1}(c)).

\subsection{Two-up-two-down collinear structure at $h=0$}
\label{Onefourth}

The theory above remains valid also if the collinear order is realized at $h=0$ in which spins are arranged in some direction in two-up-two-down manner $\uparrow \uparrow \downarrow \downarrow$ (the so-called $1/4$-structure). It appears, for instance, in the model considered in Sec.~\ref{Model} at large enough $D-E$ and $J_2>J_1/2$ (as it is seen from Eqs.~\eqref{AJq} and \eqref{Ak}, AF ordering discussed above appears at $J_1 > 2 J_2$). All the results of Sec.~\ref{Theor} are applicable in this case if one defines $\mathbf{k}_0$ as the vector of the $1/4$-structure. In particular, the $1/4$-structure is given in the classical spin chain as $S_j=S\sqrt2\cos(k_0R_j+\pi/4)$, where $k_0=\pi/2a$ and $a$ is the lattice spacing.

Our theory can analytically describe some results obtained in model \eqref{ham1} in Ref.~\cite{zh} in the framework of real-space mean-field approach. A complicated magnetic phase diagram of MnWO$_4$ observed experimentally~\cite{arkenbout2006,nojiri2011,mitamura2012} was qualitatively reproduced in Ref.~\cite{zh} with parameters
\begin{equation}
\label{Zhpar}
  J_1=1, \quad J_2 = 2, \quad D=0.4, \quad E=0.2, \quad S=5/2
\end{equation}
which yield $\mathbf{k} \approx 0.54 \pi$, $J_{\mathbf{k}}-J_{\mathbf{k}_0} \approx 0.125$, and $J_{\mathbf{k}}-J_{3\mathbf{k}} \approx 1.94$.

We find that if the field is directed along the easy axis, transitions take place from the $1/4$-structure to the YZ state and then to the CAF phase when the field increases. In this case, the CAF state consists of four magnetic sublattices forming two pairs. Within each pair, spins are oriented in the same direction. Spins from different pairs are oriented as in the CAF state of conventional antiferromagnet. Eqs.~\eqref{h1} and \eqref{h2} give $h_1 \approx 2.7$ and $h_2 \approx 7.4$ with parameters \eqref{Zhpar}. Corresponding numerical results of Ref.~\cite{zh} are approximately $1.4$ and $6.9$. The discrepancy in $h_1$ is attributed to its rather small value, which shows an importance of higher order terms in $D$ and $E$ neglected above. Taking into account the second order term (see Eq.~(23) of Ref.~\cite{utesov2018}), we obtain $h_1 \approx 2.3$ in better agreement with Ref.~\cite{zh}. Thus, our theory satisfactorily describes the numerics in this case.


\section{Summary and conclusions}
\label{Conc}

To conclude, we discuss different scenarios of phase transitions in frustrated antiferromagnets with biaxial anisotropy or dipolar forces in magnetic field applied along the easy axis. The magnetic field is assumed to be not very close to the saturation field. There are well known scenarios of phase transitions shown in Fig.~\ref{fig1}(a) and \ref{fig1}(b): the conventional spin-flop transition and the flop of the spiral plane at strong and weak easy-axis anisotropy, respectively. We demonstrate that much less studied scenarios can appear at moderate anisotropy which are presented in Figs.~\ref{fig1}(c)--(e) and in which magnetic field induces first-order transitions to spiral phases from the collinear one. In particular, the sequence of phase transitions shown in Fig.~\ref{fig1}(c) can be interpreted as a splitting of the spin-flop transition shown in Fig.~\ref{fig1}(a) into two transitions with the intermediate spiral phase. Critical fields of these transitions are given in the mean-field approximation by Eqs.~\eqref{h1} and \eqref{h2}, by Eqs.~\eqref{h1} and \eqref{hsp}, and by Eq.~\eqref{hsp} for scenarios shown in Figs.~\ref{fig1}(c), \ref{fig1}(d), and \ref{fig1}(e), respectively. Corresponding necessary conditions for realization of these scenarios are given by Eqs.~\eqref{cond1}--\eqref{cond3} and \eqref{exyz}; Eqs.~\eqref{cond1}, \eqref{cond2}, \eqref{cond4}, and \eqref{cond5}; and Eqs.~\eqref{cond1}, \eqref{cond2}, \eqref{cond4}, and \eqref{cond6}.

We demonstrate both analytically and numerically (using Monte-Carlo simulations) the appearance of scenarios shown in Figs.~\ref{fig1}(c)--(e) in particular anisotropic Heisenberg models with competing exchange couplings. We show also that the sequence of phase transitions presented in Fig.~\ref{fig1}(c) was found in MnWO$_4$ both experimentally \cite{arkenbout2006,nojiri2011,mitamura2012} and numerically \cite{zh} (in the relevant model) and our theory reproduces the numerical findings even quantitatively.

It should be noted that some of the scenarios found above in the frustrated antiferromagnets can appear also in anisotropic systems with a monoaxial Dzyaloshinskii-Moriya interaction (which produces a spiral ordering at sufficiently weak anisotropy). In particular, we have found using Monte-Carlo simulations that the scenario shown in Fig.~\ref{fig1}(c) can arise. A detailed discussion of these models will be reported elsewhere.

\begin{acknowledgments}

We thank A.O. Sorokin for valuable discussion. The reported study was funded by RFBR according to the research project 18-02-00706.

\end{acknowledgments}


\bibliography{TAFbib}

\end{document}